\documentclass[aps,pra,showpacs,twocolumn,superscriptaddress]{revtex4}
\usepackage{array}
\usepackage{booktabs}
\usepackage{tabu}
\usepackage{dcolumn}
\usepackage{amsmath}
\usepackage{amsfonts}
\usepackage{float}
\usepackage{amssymb}
\usepackage{graphicx,color}
\usepackage[colorlinks={true}]{hyperref}
\usepackage{graphicx}
\usepackage{subfigure}
\usepackage{graphicx}% Include figure files
\usepackage{dcolumn}% Align table columns on decimal point
\usepackage{bm}% bold math

\def\be{\begin{equation}}
  \def\ee{\end{equation}}
\def\bea{\begin{eqnarray}}
\def\eea{\end{eqnarray}}
\def\f{\frac}
\def\n{\nonumber}
\def\l{\label}
\def\p{\phi}
\def\o{\over}
\def\R{\rho}
\def\pa{\partial}
\def\om{\omega}
\def\na{\nabla}
\def\P{\Phi}
\begin{document}

\title{Quantum speed limit time for the damped Jaynes-Cummings and Ohmic-like dephasing models in Schwarzschild space-time}

\author{S. Haseli}
\email{soroush.haseli@uut.ac.ir}
\affiliation{Faculty of Physics, Urmia University of Technology, Urmia, Iran.}

\date{\today}% It is always \today, today,

\def\be{\begin{equation}}
  \def\ee{\end{equation}}
\def\bea{\begin{eqnarray}}
\def\eea{\end{eqnarray}}
\def\f{\frac}
\def\n{\nonumber}
\def\l{\label}
\def\p{\phi}
\def\o{\over}
\def\R{\rho}
\def\pa{\partial}
\def\om{\omega}
\def\na{\nabla}
\def\P{\Phi}
%\nofiles
\begin{abstract}
Quantum theory sets the bound on the minimal evolution time between initial and final states of the quantum system.  This minimal evolution time  can be used to specify the maximal speed of the evolution in open and closed quantum systems. Quantum speed limit is one of the interesting issue in the theory of open quantum systems.  One may investigate the influence of the relativistic effect on the quantum speed limit time. When several observers are placed in different inertial or non-inertial frames, or in Schwarzschild space-time, the relativistic effect should be taken into account.  In this work, the quantum speed limit time in Schwarzschild space-time will be studied for two various model consist of damped Jaynes-Cummings and Ohmic-like dephasing. First, it will be observed that how quantum coherence is affected by Hawking radiation. According to the dependence of  quantum speed limit time on quantum coherence and the dependence of quantum coherence on  relative distance of quantum system to event horizon $R_{0}$, it will be represented that the quantum speed limit time in Schwarzschild space-time is decreased by increasing $R_{0}$ for damped Jaynes-Cummings model and conversely, It is increased by increasing $R_{0}$ for Ohmic-like dephasing model . 
\end{abstract}
\maketitle
%=============================================================%
%=============================================================%
%============== Abstract =======================================%
%=============================================================%
%=============================================================%

\section{Introduction}\label{intro}
The minimum time for the evolution of a quantum system from an initial state at time $\tau$ to target  state at time $\tau + \tau_D $ is known as quantum speed limit QSL-time  $\tau_{(QSL)}$, where $\tau_D$ is driving time. QSL-time can be interpreted as a generalization of the time-energy uncertainty principle. It determines the maximum speed of a quantum evoulution. Quantum speed limit time is used for many topics in quantum information theory such as, quantum communication \cite{Bekenstein}, exploration of accurate bounds in quantum metrology \cite{Giovanetti}, computational bounds of physical systems \cite{Lloyd} and quantum optimal control algorithms \cite{Caneva}.

For closed quantum systems, which are isolated from its surroundings, the dynamic is unitary and different bounds on QSL-time have been obtained based on Bures angle and relative purity as the distance measures between initial and target state \cite{Uhlmann,Pfeifer,Giovannetti,Pfeifer1,Chau,Deffner,Mandelstam,Margolus}.  Mandelstam-Tamm (MT) bound \cite{Mandelstam} and Margolus-Levitin (ML) bound \cite{Margolus} are the most famous bounds of QSL-time for closed quantum systems . Due to inescapable interaction between the real quantum system and its environment the study of the open quantum systems is an important topic in quantum information theory \cite{Davies,Alicki,Breuer}. So, it is of particular importance to calculated the QSL-time for open quantum systems.  By making use of various distance measures, different bounds on the QSL-time  have been proposed for open quantum systems \cite{Campo,Carlini,Brody,Taddei,Deffner1,Xu,Liu,Zhang,Campaioli,Wu1,Sun}. In general, there exist two types of QSL-time, Mandelstam-Tamm bound and Margolus-Levitin bound. In Refs. \cite{Giovannetti,Pfeifer1}, the authors have provided the generalization of the (MT) and (ML) bounds for nonorthogonal states and for driven systems. Deffner et al. formulated the unified bound of QSL-time including both (MT) and (ML) types for non-Markovian dynamics \cite{Deffner1}. In recent years, QSL-time has been studied from different perspectives such as the effect of the decoherence on QSL-time \cite{Mukherjee,Marvian,Dehdashti,Brouzos}, the role of the initial state on QSL-time \cite{Wu} and the applications of the QSL-time in quantum phase transition\cite{Wei}.

It is worth noting that QSL-time $\tau_{(QSL)}$ can be interpreted as the potential capacity for further evolution acceleration. If $\tau_{(QSL)}=\tau_{D}$ then the evolution is now in the situation with the highest speed, thus the evolution has not the potential capacity for further acceleration. However, when $\tau_{(QSL)} < \tau_D$, the potential capacity for further acceleration will be greater. Another important point to be noted here is that, when the coupling strength  between the system and environment is weak $\tau_{(QSL)}$ tends to the actual driving time $\tau_{D}$. On the contrary, in the  strong coupling  between the system and environment, $\tau_{(QSL)}$ can be reduce below the actual driving time $\tau_D$ \cite{Deffner1}.  In the other words, strong coupling can increase the speed of the quantum evolution while the week system-environment couplings can not increase  the speed of the quantum evolution. In this paper,  relative purity is used as the distance measure to derive a QSL-time  for open system dynamics. The advantage of using relative purity is that it is applicable for both mixed and pure initial states\cite{Zhang}. 

In Ref. \cite{Zhang}, Zhang et al. have shown that QSL-time depend on quantum coherence. They have shown that in the damped Jaynes-Cummings model, QSL-time decreases by increasing the quantum coherence while in Ohmic-like dephasing model QSL-time increase by increasing quantum coherence. It should be noted that the quantum coherence can  also be  affected by Hawking radiation in curved space-time \cite{Hosler,Ahn}. It can be shown that the quantum coherence of quantum system would degrade when quantum system gets close to  the event horizon. Thus, it is logical to expect that the QSL-time can be affected by Hawking radiation in curved space-time. It is interesting to investigate how Hawking radiation would affect on the QSL-time. To demonstrate this, the simplest black hole is used : Schwarzchild  black
hole with Dirac field states. Here, the Dirac field state is  considered  instead of bosonic state
because due to Pauli’s exclusion principle  there exist utmost one particle for each spin in one mode. In this work we will consider the setting in which quantum system freely falls into the Schwarzschild black hole then hovers near the event horizon and finally it interacts with surroundings in the damped Jaynes-Cummings or Ohmic like dephasing models.

This work is organized as follows. In Sec. \ref{sec:2}, we will discuss about QSL-time for open quantum systems which is introduced based on relative purity. In Sec. \ref{sec:3}, we discuss about important exclusivity of Dirac fields in Schwarzschild space-time to express the QSL-time for  open quantum systems in the background of a black hole. In Sec. \ref{sec:4} ,  the QSL-time of a single qubit system is investigated in the damped Jaynes-Cummings and Ohmic like dephasing models in Schwarzschild space-time . The conclusion of this work is given in Sec.\ref{sec:5}.
\section{Quantum speed limit time based on relative purity}\label{sec:2}
In this section the dynamics of the open quantum systems is considered .  The state of the system at time $t$ is described by density matrix $\rho_t$. The evolution of such a quantum system can be described by the time-dependent nonunitary equations of the form $\dot{\rho}_{t}=\mathcal{L}_{t}(\rho_{t})$, where $\mathcal{L}_{t}$ is the positive generator  \cite{Breuer}.  Here we want to calculate the minimum time which is necessary to evolve from the state at time $\tau$  to target state at time $\tau + \tau_D$, where $\tau_D$ is the driving time of the open quantum system. In order to obtain this minimal time one should choose an appropriate distance measure to define the QSL-time. In Refs. \cite{Zhang,Campaioli,Wu1}, authors have introduced the QSL-time based on relative purity. The advantage of their QSL-time is that, it is applicable for both mixed and pure initial states. One can  define the relative purity $f(\tau)$ between initial state $\rho_{\tau}$ and target state $\rho_{\tau+\tau_D}$ as  \cite{Audenaert}
\begin{equation}\label{relative purity}
f(\tau + \tau_D)=\frac{tr(\rho_{\tau}\rho_{\tau + \tau_D})}{tr(\rho_{\tau}^{2})}.
\end{equation}
Using the method outlined in Ref. \cite{Zhang}, one can derive the (ML) bound of QSL-time  for non-unitary  dynamics as 
\begin{equation}\label{ML}
\tau \geq \frac{\vert f( \tau + \tau_D ) -1 \vert tr (\rho_{\tau}^{2})}{\overline{ \sum_{i=1}^{n} \sigma_{i} \rho_{i}}},
\end{equation}
where $\sigma_{i}$ and $\rho_{i}$ are the singular values of $\mathcal{L}_{t}(\rho_{t})$ and $\rho_{\tau}$, respectively and  $\overline{\mathcal{A}}=\frac{1}{\tau_{D}} \int_{\tau}^{\tau + \tau_{D}} \mathcal{A} dt$. In a similar way,  (MT) bound of QSL-time for non-unitary evolution can be obtain as 
\begin{equation}\label{MT}
\tau \geq \frac{\vert f( \tau + \tau_D ) -1 \vert tr (\rho_{\tau}^{2})}{\overline{ \sum_{i=1}^{n} \sigma_{i}^{2}}}.
\end{equation}
From Eqs. (\ref{ML}) and (\ref{MT}), unified bound can be formulated as follows
\begin{equation}\label{(QSL)T}
\tau_{(QSL)}=\max \lbrace \frac{1}{\overline{ \sum_{i=1}^{n} \sigma_{i} \rho_{i}}}, \frac{1}{\overline{ \sqrt{\sum_{i=1}^{n} \sigma_{i}^{2}}}} \rbrace \times \vert f( \tau + \tau_D ) -1 \vert tr (\rho_{\tau}^{2}).
\end{equation}
Zhang et al. have shown that this QSL-time is depend on the coherence of initial state $\rho_{\tau}$ \cite{Zhang}. They have also shown that the (ML) type bound of the QSL-time  in Eq. \ref{ML} is tight for the open quantum systems.
\section{Dirac fields in Schwarzschild space-time}\label{sec:3}
\begin{figure}[t]
\center
  \includegraphics[width=0.45\textwidth]{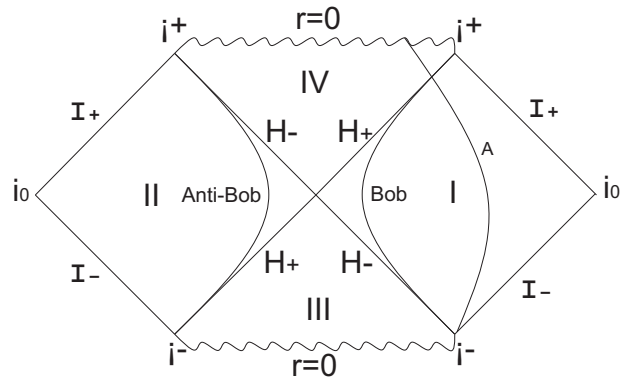}
\caption{The Penrose diagram of Schwarzschild space-time which
shows the world-line of Alice, Bob and Anti-Bob. $i_0$ denotes the spatial
infinities, $i^{−}$ ($i^+$) represents time-like past (future) infinity. $I_{-}$ ($I_{+}$)
shows light-like past (future) infinity. $H_{\pm}$ show the event horizons of
the black hole. }
\label{blackhole}       % Give a unique label
\end{figure}
In this section,  we discuss over important exclusivity of Dirac fields in Schwarzschild space-time to express the QSL-time for non-unitary open quantum dynamic in the background of a black hole. Let's consider a Schwarzschild space-time which is given by the metric in the form below 
\begin{equation}\label{metric}
ds^{2}=-\left( 1-\frac{2M}{r} \right) dt^{2} + \left( 1-\frac{2M}{r} \right)^{-1}dr^{2} + r^{2}d \Omega^{2},
\end{equation}  
where $M$ and $r$ are the mass and radius of the black hole, respectively. $d \Omega^{2} = d \theta^{2} + \sin^{2}\theta d \phi^{2} $, is the line element in the unit sphere. It is worth noting that, near the event horizon $r_h$, this metric   has similar structure with Rindler horizon in flat
space-time \cite{Martin}. One can see the Penrose diagram of Schwarzschild space-time in Fig. \ref{blackhole}.

Let's  consider the massless Dirac equation in the curved space-time as \cite{Birrell}
\begin{eqnarray}\label{Dirac11}
&&\frac{-\gamma_{0}}{\sqrt{1-\frac{2M}{r}}} \frac{\partial \psi}{\partial t} + \nonumber \\
&&+\gamma_{1} \sqrt{1-\frac{2M}{r}} \left( \frac{\partial}{\partial r} + \frac{1}{r} + \frac{M}{2r(r-2M)}  \right)\psi +   \nonumber \\
 &&+\frac{\gamma_{2}}{r} \left(\frac{\partial}{\partial \theta} + \frac{1}{2} \cot \theta \right)\psi + \frac{\gamma_{3}}{r \sin \theta}\frac{\partial \psi}{\partial \phi} =0,
\end{eqnarray}
where $\gamma_{i}$'s $(i=0,1,2,3)$ are the 4 by 4 Dirac matrices. Solving masless Dirac equation (\ref{Dirac11}) near the event horizon $r_h$ leads to positive (fermions) frequency outgoing solutions outside and inside regions of the event horizon as   \cite{Brill,Brill1}
\begin{equation}
\psi_{k}=\left\{ \begin{array}{cl}
 \xi e^{-i\omega u} & \qquad r > r_{h}\\ \\
 \xi e^{i\omega u} & \qquad r < r_{h}\end{array},\right.
\end{equation}
where $u=t-r^{*}$ and $r^{*}=r + 2M \ln \vert \frac{2M-r}{2M} \vert$ is the  tortoise coordinate. The goal is to determine the vacuum structure for different observer. For this, Let's introduce the light-like Kruskal coordinates  as \cite{Martin}
\begin{equation}\label{Kruskal}
\mathcal{U} = - \frac{1}{k}\exp \left[ -k(t-r^{*}) \right] , \quad \mathcal{V} = - \frac{1}{k}\exp \left[ k(t+r^{*}) \right],
\end{equation} 
where $k=1/4M$ is the surface gravity. With regard to light-like Kruskal coordinates, the Schwarzschild
metric has the following form \cite{Pan,Wang}
\begin{equation}
ds^{2}=-\frac{1}{2kr}e^{-2kr}d\mathcal{U}d\mathcal{V} + r^{2}d \Omega^{2}.
\end{equation}
There exist three regions with different physical time-like vectors \cite{Feng}. In first region, for time-like vector in the form $\partial_{\hat{t}} \propto (\partial_{\mathcal{U}} + \partial_\mathcal{V})$, specific parameter $t$ is associated with the proper time of a free-falling observer (Alice) near the horizon. This time-like vector is similar to the Minkowskian time-like Killing vector and the structure of the Hartle–Hawking vacuum $\vert 0_H \rangle $ is similar  to the structure of the Minkowski vacuum $\vert 0_M \rangle $. In second region, $\partial_{t} \propto ( \mathcal{U} \partial_{\mathcal{U}} + \mathcal{V} \partial_\mathcal{V})$ is the time-like Killing vector, which is  related to an observer (Bob) with proper acceleration $a=k/\sqrt{1-\frac{2M}{r_{0}}}$, where $r_{0}$ is the position of the observer with respect to event horizon, it is assumed that this distance is small enough. $\partial_t$ is similar to Rindler vacuum in flat space.  Note that the Rindler approximation is only logical when $R_{0}  -1 <<1$, where $R_{0}=r_0/r_h=r_{0}/2M$ \cite{Martin}. For this time-like Killing vector, the vacuum structure is known as the Boulware vacuum $\vert 0 \rangle_{I}$.  $-\partial_t$ is another Killing vector. $-\partial_t$ enables us to introduce another Boulware vacuum which is known as Anti-Boulware vacuum $\vert 0 \rangle_{II}$.

The Hartle–Hawking vacuum structure is made from   a variety of frequency modes as $\vert 0_H \rangle = \otimes \vert 0_{\omega_{i}} \rangle_H$, in a similar way for first excitation $ \vert 1_H \rangle = \otimes \vert 1_{\omega_{i}} \rangle_ H $. One can define the Hartle-Hawking vacuum $\vert 0_{\omega_{i}} \rangle$ and its first excitation $\vert 1_{\omega_{i}} \rangle$ as \cite{Martin,Alsing}
\begin{eqnarray}\label{Dirac}
\vert 0_{\omega_{i}} \rangle_{H}&=&\frac{1}{\sqrt{1+e^{-\Omega \sqrt{1-\frac{1}{R_{0}}}}}} \vert 0_{\omega_{i}} \rangle_{I}  \vert 0_{\omega_{i}} \rangle_{II} 
 \nonumber \\
&+&\frac{1}{\sqrt{1+e^{\Omega \sqrt{1-\frac{1}{R_{0}}}}}} \vert 1_{\omega_{i}} \rangle_{I}  \vert 1_{\omega_{i}} \rangle_{II}, \\
\vert 1_{\omega_{i}} \rangle_{H}&=&\vert 1_{\omega_{i}} \rangle_{I}  \vert 0_{\omega_{i}} \rangle_{II},
\end{eqnarray}
where $\Omega =\frac{\omega}{T_{H}} =\frac{2 \pi \omega}{k}$ is the mode frequency measured by Bob and $T_{H}=k/2\pi$ is the Hawking temperature. This formulation enables us to investigate the  QSL-time in  Schwarzschild space-time.
\section{Quantum speed limit time of the dynamics in  Schwarzschild space-time }\label{sec:4}
In this section we investigate the effect of Hawking radiation on QSL-time $\tau_{(QSL)}$. First, let's assume that the quantum system $B$ is in possession of Bob in single-qubit state  $\rho_{0}^{B} = \frac{1}{2} (\mathcal{I} + \sum_{i=1}^{3} r_{i} \sigma_{i})$,  where  $\mathcal{I}$ is the identity operator , $\sigma_{i}$'s ($i=1,2,3$) are the Pauli operators, and $r_{i}$'s are the components of Bloch vector. Bob falls toward the black hole and then locates at a fixed distance $r_0$ outside the event horizon. 

\begin{figure}[t]
  \includegraphics[width=0.5\textwidth]{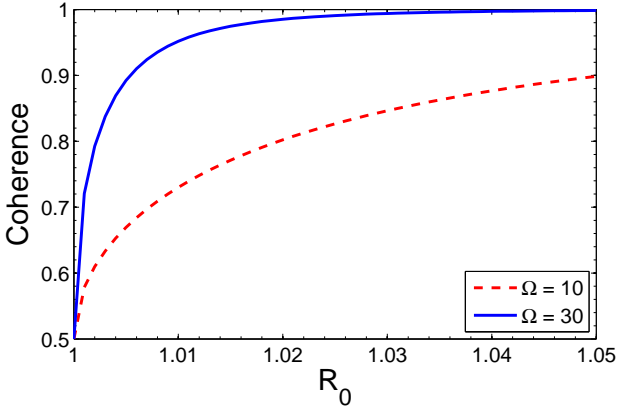}
\caption{The quantum coherence  as function of $R_0$. The maximally coherent state  with $C(\rho_{0}^{B})=1$ has been chosen as an initial state of fall free particle. The relative distance of Bob to event horizon  is $R_0 \leq 1.05$, so Rindler approximation can be hold. }
\label{fig:1}       % Give a unique label
\end{figure}

Let us assume that Bob has a detector which only detects mode with frequency $\omega$. Thus, the states associate with mode $\omega$ must be specified in Boulware basis. From the viewpointe of Bob, under the single-mode approximation, Hartle-Hawking vacuum and its excitation in Boulware basis is rewritten as 
\begin{eqnarray}\label{Dirac1}
\vert 0_{\omega} \rangle_{H}&=&\jmath_{-}  \vert 0_{\omega} \rangle_{I}  \vert 0_{\omega} \rangle_{II} +\jmath_{+} \vert 1_{\omega} \rangle_{I}  \vert 1_{\omega} \rangle_{II}, \\ \nonumber
\vert 1_{\omega} \rangle_{H}&=&\vert 1_{\omega} \rangle_{I}  \vert 0_{\omega} \rangle_{II},
\end{eqnarray}
where $\jmath_{+}=[1+\exp(\Omega \sqrt{1-\frac{1}{R_{0}}})]^{-\frac{1}{2}}$ and $\jmath_{-} =[1+\exp(-\Omega \sqrt{1-\frac{1}{R_{0}}})]^{-\frac{1}{2}}$. The subscripts $I$ and $II$ show the states associated to the Rindler region $I$ and region $II$ respectively. In particular, after transforming Bob’s states according to Eq. (\ref{Dirac1}) and by tracing over the degrees  in region $II$, one can obtain the density matrix for single-qubit system under the effect of Hawking radiation as 
\begin{equation}\label{hawkingstate}
\rho_{I0}^{B}=\frac{1}{2}\left(
\begin{array}{cc}
 \jmath_{-}^{2}(1+r_3) & \jmath_{-}(r_{1}-ir_2) \\
 \jmath_{-}(r_{1}+ir_2) & (1-r_3)+\jmath_{+}^{2}(1+r_3) \\
\end{array}
\right).
\end{equation}
When Bob falls toward the black hole and locates at a fixed distance $r_0$ outside the event horizon then one can find the coherence of the Bob's state as $C(\rho_{I0}^{B})=\jmath_{-}\sqrt{r_{1}^{2}+r_{2}^{2}}$. Due to the mathematical form of the coefficient $\jmath_{-}$ one can see the quantum coherence decreases when the distance of Bob from event horizon, i.e. $R_0$ decreases. In Fig. \ref{fig:1} the quantum coherence is plotted as a function of $R_{0}$ for the state with initial coherence $C(\rho_{0}^{B})=\sqrt{r_{1}^{2}+r_{2}^{2}}=1$ . As can be seen from Fig. \ref{fig:1}, the quantum coherence decreases when the distance of Bob from event horizon $R_0$ decreases. Also, one can see the quantum coherence increases strongly when the the mode frequency increases.  In other words quantum coherence increases when Hawking temperature $T_H$  decreases.

In our scenario the quantum system $B$ is located  at a fixed distance $r_0$ outside the event horizonis and it is in a direct interaction with the environment. We consider two types of decoherence model to investigate the effect of Hawking radiation on QSL-time : damped Jaynes-Cummings and Ohmic-like dephasing models. In Ref. \cite{Zhang}, a simple structure for these model is used in the presence of the relativistic effects. So, using the ordinary structure of these two interaction model in Schwarzschild black hole is applicable and logical due to considering Dirac field state  instead of bosonic state and applying single-mode aproximation. Actually, in must studies \cite{Zhang,Wangmaster,Yu,Ramzan,Ramzan1,Ramzan2} the authors use ordinary structure for the generator of the dynamics of the system in Damped Jaynes-Cummings and Ohmic-like dephasing models.
\subsection{Damped Jaynes-Cummings}
Let us consider the exactly solvable damped Jaynes-Cummings model for a two-level system  which is coupled to a leaky single mode cavity \cite{Breuer1,Garraway}. The environment is assumed to be initially in a vacuum state. The non-unitary  positive generator of the dynamics of the system is given by
\begin{equation}\label{master1}
\mathcal{L}_{t}(\rho_{t})=\gamma_{t}(\sigma_{-}\rho_{t}\sigma_{+}-\frac{1}{2}\sigma_{+}\sigma_{-}\rho_{t}-\frac{1}{2}\rho_{t}\sigma_{+}\sigma_{-}),
\end{equation}
where $\sigma_{\pm} = \sigma_{1} \pm i \sigma_{2}$ and $\gamma_{t}$ is the
time-dependent decay rate. If  there exist one excitation in
the compound atom-cavity system then the environment can be described by an effective Lorentzian spectral density as
\begin{equation}\label{so}
J(\omega)=\frac{1}{2 \pi}\frac{\gamma_{0}\lambda}{(\omega_{0}-\omega)^{2}+\lambda^{2}},
\end{equation} 
where $ \omega_{0}  $ is the frequency of the single-qubit system, $\lambda$ denotes the spectral width and $\gamma_{0}$ represents the coupling strength.  In Ref. \cite{Zhang}, authors use a simple structure for this model in the presence of the relativistic effects. So, using the ordinary structure of this interaction model  in Schwarzschild black hole is applicable and logical due to considering Dirac field state  instead of bosonic state and applying single-mode aproximation. So, Eqs. \ref{master1} are still applicable in the Schwarzschild spacetime.

\begin{figure}[t]
  \includegraphics[width=0.5\textwidth]{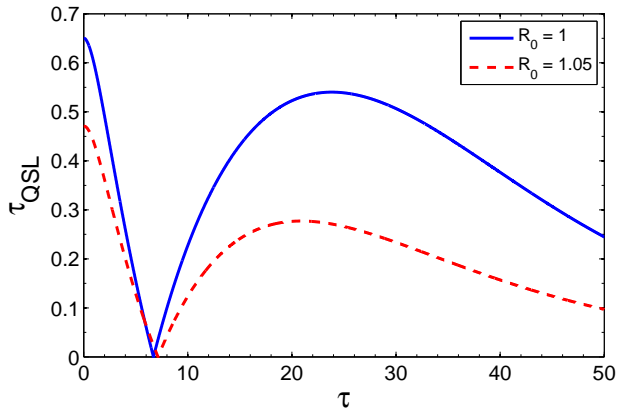}
\caption{The QSL-time  for the damped Jaynes-Cummings model as a
function of the initial time parameter $\tau$. The coupling is weak i.e. $\gamma_{0}=0.1 \lambda$ and $\Omega=10$. }
\label{fig:2}       % Give a unique label
\end{figure}

\begin{figure}[t]
  \includegraphics[width=0.5\textwidth]{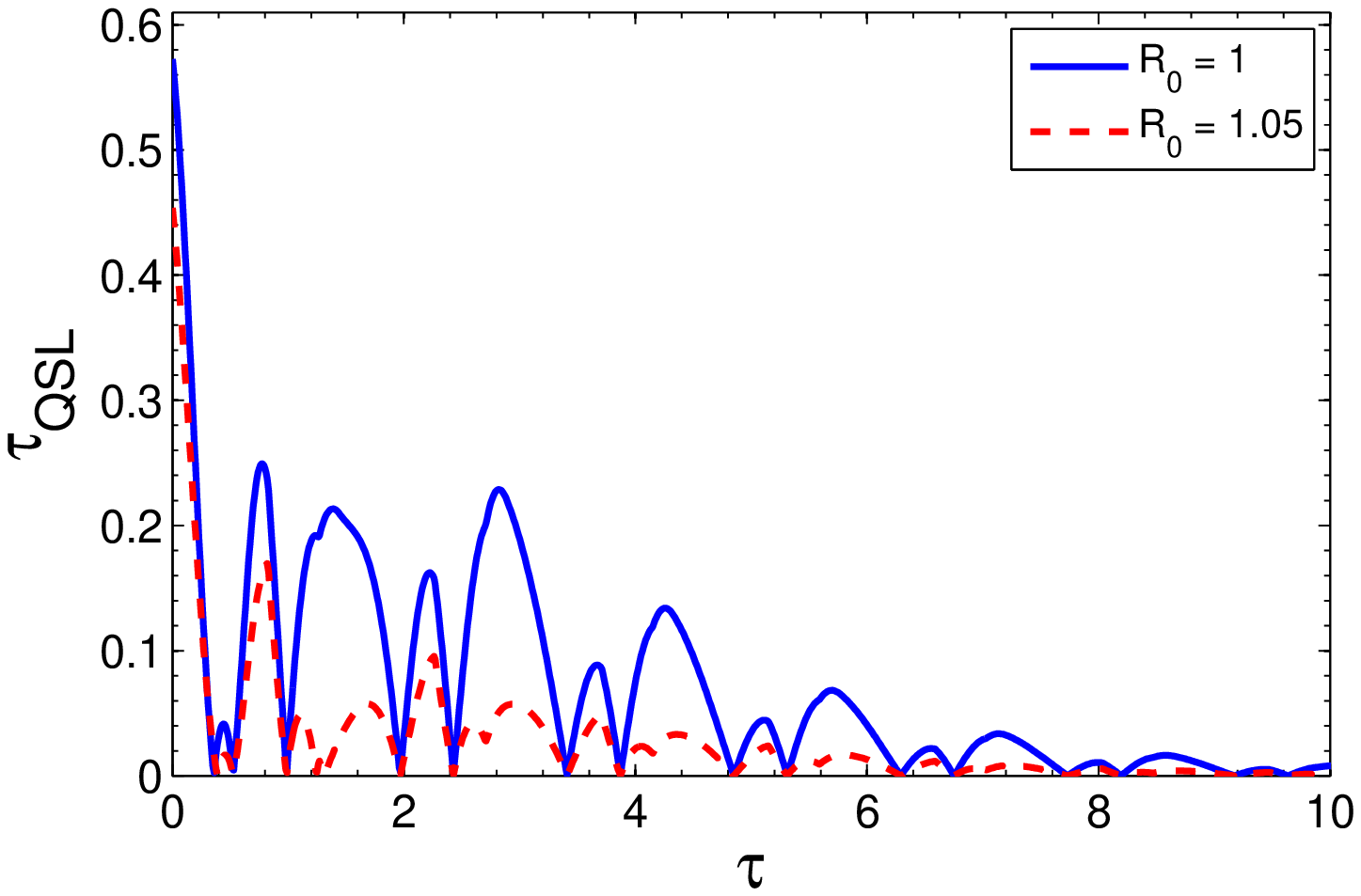}
\caption{The QSL-time  for the damped Jaynes-Cummings model as a
function of the initial time parameter $\tau$. The coupling is weak i.e. $\gamma_{0}=10 \lambda$ and $\Omega=10$. }
\label{fig:3}       % Give a unique label
\end{figure}

The time-dependent decay rate can be written as 
\begin{equation}
\gamma_{t}=\frac{2 \gamma_{0}\lambda \sinh (dt/2)}{d \cosh(dt/2)+\lambda \sinh(dt/2)},
\end{equation}
where $d=\sqrt{\lambda^{2}-2\gamma_{0}\lambda}$. When the coupling is weak i.e. $\lambda > 2 \gamma_{0}$, the dynamics is Markovian and information flow to the environment in irreversible manner. In contrast, when we have the strong coupling between system and environment i.e. $\lambda < 2 \gamma_{0} $ the dynamics is non-Markovian and information flow back to the system from environment\cite{Breuerx}. Near the event horizon,  the  density operator $\rho_{tI}^{B}$ of the open quantum system is obtained as 
\begin{equation}
\rho_{tI}^{B}=\left(
\begin{array}{cc}
 1-p(t) (1-\rho _{11}) & \sqrt{p_t} \rho _{12} \\
 \sqrt{p_t} \rho _{21} & p_t \rho _{22} \\
\end{array}
\right) ,
\end{equation}
where 
\begin{eqnarray}\label{element}
\rho_{11}&=&\frac{1}{2}(\jmath_{-}^{2}(1+r_3)), \quad \rho_{22}=\frac{1}{2}((1-r_3)+\jmath_{+}^{2}(1+r_3)), \nonumber \\
\rho_{12}&=&\frac{1}{2}(\jmath_{-}(r_{1}-ir_2)), \quad \rho_{21}=\frac{1}{2}(\jmath_{-}(r_{1}+ir_2)),
\end{eqnarray}
with $p_t=\exp(-\int_{0}^{t} \gamma_{t^{\prime}} dt^{\prime})$. In Ref. \cite{Zhang}, authors show that the (ML) type bound of
the QSL-time in Eq. \ref{ML} is tight for the open quantum systems. Considering this fact, one can obtain QSL-time for damped Jaynes-Cummings in Schwarzschild space-time as
\begin{equation}\label{(QSL)Tj}
\tau_{(QSL)}= \frac{1}{\overline{  \sigma_{1} \rho_{1}+\sigma_{2} \rho_{2}}}  \times \vert f( \tau + \tau_D ) -1 \vert tr (\rho_{\tau}^{2}),
\end{equation}
where $\vert f( \tau + \tau_D ) -1 \vert tr (\rho_{\tau}^{2})=\frac{1}{2} \vert \jmath_{-}\sqrt{p_{\tau}p_{\tau + \tau_{D}}}C(\rho_{0}^{B})^{2}-p_{\tau+\tau_{D}}(1+\jmath_{+}^{2}-\jmath_{-}^{2}r_{3})-p_{\tau}^{2}(1+\jmath_{+}^{2}-\jmath_{-}^{2}r_{3})^{2} + p_{\tau}(1+\jmath_{+}^{2}-\jmath_{-}^{2}C(\rho_{0}^{B})^{2}-\jmath_{-}^{2}r_3+p_{\tau+\tau_{D}}(1+\jmath_{+}^{2}-\jmath_{-}^{2}r_{3})^{2})\vert$ and $\sigma_{1} \rho_{1}+\sigma_{2} \rho_{2} = \frac{\vert \dot{p}_{t} \vert}{4}\sqrt{(2 \jmath_{-}^{2}(1+r_3)-4)^{2}+\jmath_{-}^{2}C(\rho_{0}^{B})^{2}/p_{t}}$. It can be seen $\tau_{(QSL)}$ is not only relate to
$r_{3}$ but also to the coherence of the initial state $C(\rho_{0}^{B})$.
  Here after,  we choose an initial state with maximum quantum coherence i.e.  $C(\rho_{0}^{B})=1$.   In Fig.(\ref{fig:2}) the QSL-time is plotted  as a function   of the initial time parameter $\tau$ in the weak coupling regime $\gamma_{0} =  0.1 \lambda$. As can be seen QSL-time is decreased by increasing the distance $R_{0}$ between Bob and  the event horizon. Fig. (\ref{fig:3}) represents the QSL-time as a function   of the initial time parameter $\tau$  in strong coupling regime. Fig. (\ref{fig:3}) shows in strong coupling the QSL-time is shorter than weak coupling. As can be seen, QSL-time is decreased by increasing $R_{0}$.

\begin{figure}[t]
  \includegraphics[width=0.5\textwidth]{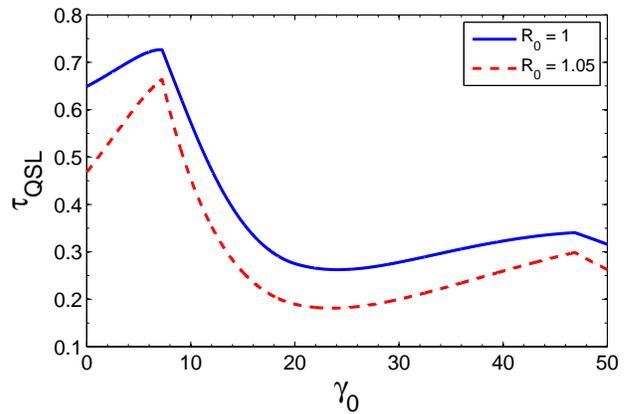}
\caption{The QSL-time  for the damped Jaynes-Cummings model as a
function of the coupling strenght $\gamma_{0}$. The value of driving time is $\tau_{D}=1$, initial time $\tau=0$ and $\Omega=10$. }
\label{fig:4}       % Give a unique label
\end{figure}

\begin{figure}[t]
  \includegraphics[width=0.5\textwidth]{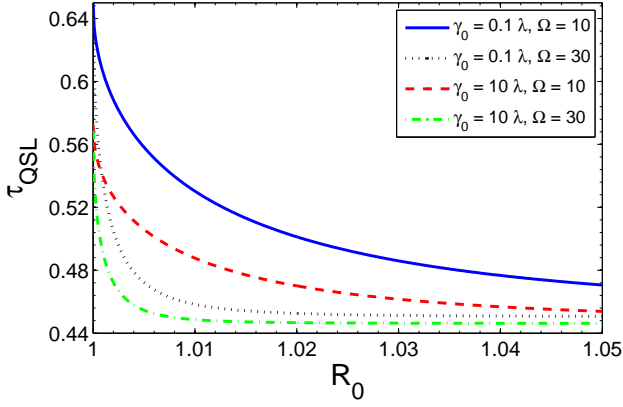}
\caption{The QSL-time  for the damped Jaynes-Cummings model as a
function of the  $R_{0}$. The value of driving time is $\tau_{D}=1$ and initial time $\tau=0$. }
\label{fig:5}       % Give a unique label
\end{figure}

In order to show the effects of  coupling strenght on QSL-time, the  QSL-time is plotted as a function of $\gamma_{0}$ in Fig. (\ref{fig:4}) for the constant driving time $\tau_D=1$. As can be seen, for both strong and weak coupling regime the QSL-time decrease when $R_{0}$ increases. In order to show the effect of the the mode frequency $\Omega$ (Hawking temperature $T_H$)  and $R_0$ on QSL-time we plot the QSL-time as a function of $R_{0}$ in strong and weak coupling limit for different value of $\Omega$ in Fig. (\ref{fig:5}). As can be seen for both of strong and weak coupling the QSL-time is increased by decreasing $R_{0}$. Also, this figure shows that the QSL-time is decreased by increasing $\Omega$ (decreasing $T_{H}$).

 According to the  dependence of QSL-time in  Schwarzschild space-time on  the coherence of initial state $C(\rho_{I0}^{B})=\jmath_{-}C(\rho_{0}^{B})$ and comparing Figs. (\ref{fig:2},\ref{fig:3},\ref{fig:4},\ref{fig:5}) with Fig.(\ref{fig:1}), It is observed that the QSL-time decreases with increasing quantum coherence.
In summary, the important results about QSL-time in  Schwarzschild space-time for Damped Jaynes-Cummings model  are given as follow : First, QSL-time decreases when the distance of Bob from event horizon, i.e. $R_0$ increases. In other word the speed of quantum evolution in damped Jaynes-Cummings model increases when the distance of open quantum system from event horizon increases. Second, QSL-time for non-Markovian regime (strong coupling) is shorter than Markovian regime (weak coupling). Third, QSL-time dedreases when the mode frequency $\Omega$ increases. Due to the inverse dependence of $\Omega$ on Hawking temperature $T_H$, it is right to say that QSL-time is decreased by decreasing  Hawking temperature.  
 \subsection{Ohmic-like dephasing model}
Let us consider the interaction between a two-level system and a
bosonic environment with spin-boson type Hamiltonian that represents a pure dephasing model. This model is the exactly solvable model \cite{Chin,Fanchini}. Note that there exist no correlation between system and environment. The non-unitary  positive generator of the dynamics of the system is given by 
\begin{equation}\label{master2}
\mathcal{L}_{t}(\rho_{t})=\gamma_{t}(\sigma_{3}\rho_{t}\sigma_{3}-\rho_{t}),
\end{equation} 
 where the time-dependent decay rate is written as 
 \begin{equation}
 \gamma_{t}=\int_{0}^{\infty}d \omega J(\omega) \coth(\frac{\hbar \omega }{2 K_{B}T})\frac{1-\cos \omega t}{\omega^{2}},
 \end{equation}
 where $T$ is temperature and $K_{B}$ is Boltzmann constant. Here, we consider the Ohmic-like spectral density as 
 \begin{equation}
 J(\omega)=\eta \frac{\omega^{s}}{\omega_{c}^{s-1}}\exp(-\frac{\omega}{\omega_{c}}),
\end{equation} 
where $\omega_{c}$ is the cutoff frequency and $\eta$ is a dimensionless coupling constant. The type of the environment is characterized by $S$. For $s \leq 1$, $s=1$ and $s \geq 1$ we have sub-Ohmic, Ohmic and super Ohmic environment respectively. In Ref. \cite{Zhang}, authors use a simple structure for this model in the presence of the relativistic effects. So, using the ordinary structure of this interaction model  in Schwarzschild black hole is applicable and logical due to considering Dirac field state  instead of bosonic state and applying single-mode aproximation. So, Eqs. \ref{master2} are still applicable in the Schwarzschild spacetime.

In the limit of zero temperature $T=0$, for $t > 0$ and  $s \geq 0$, the dephasing rate can be written as 
\begin{equation}
\gamma_{t}=\eta \left[ 1- \frac{\cos[(s-1)\arctan(\omega_c t)]}{(1+\omega_{c}^{2}t^{2})^{\frac{s-1}{2}}} \right]\Gamma(s-1), 
\end{equation}   
where $\Gamma(.)$ is the Euler Gamma function. Note that, in special case when $s\rightarrow 1$, the dephasing rate is obtained as $\gamma_{t}(s=1)= \eta \ln[1+\omega_{c}^{2}t^{2}]$. In this decoherence model we find the reduced density operator $\rho_{tI}^{B}$ as
\begin{equation}
\rho_{tI}^{B}=
  \left(
\begin{array}{cc}
 \rho _{11} & q_t \rho _{12} \\
  q_t \rho _{21} & \rho _{22} \\
\end{array}
\right) ,
\end{equation}
where $\rho_{ij}$, $i,j=1,2$ has introduced in Eq.\ref{element} and $q_{t}=exp[-\gamma_{t}]$. 

Similar to the case of damped Jaynes-Cummings model in Eq. \ref{(QSL)Tj}, the quantum speed limit time for Ohmic-like dephasing model in Schwarzschild space-time is obtained as 
\begin{equation}\label{(QSL)D}
\tau_{(QSL)}=\frac{\jmath_{-}C(\rho_{0}^{B})\vert q_{\tau}^{2}-q_{\tau}q_{\tau + \tau_{D}}\vert}{\frac{1}{\tau_{D}}\int_{\tau}^{\tau + \tau_{D}} \vert \dot{q}_{t}\vert dt}.
\end{equation}
As can be seen from Eq. \ref{(QSL)D}, unlike the damped Jaynes-Cummings model  $\tau_{(QSL)}$ is independent of $r_{3}$. It just depends on the dephasing rate of the Ohmic-like environment
and has a direct and linear dependence with coherence of the initial state $C(\rho_{I0}^{B})=\jmath_{-}C(\rho_{0}^{B})$  under a given driving time $\tau_D$.
\begin{figure}[t]
  \includegraphics[width=0.5\textwidth]{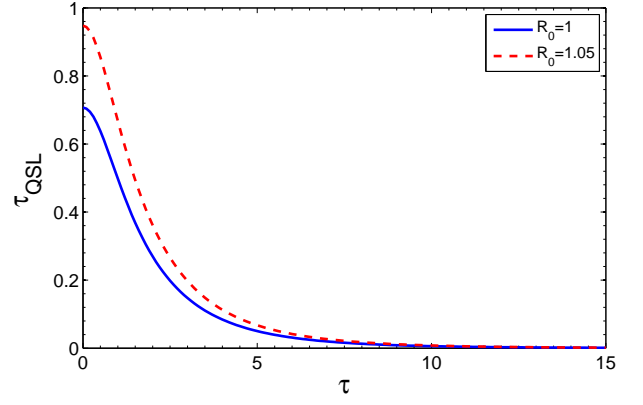}
\caption{The QSL-time  for the Ohmic-like dephasing model as a
function of the initial time parameter $\tau$ for sub-Ohmic environment. The Ohmicity parameter is $s=0.5$ and $\Omega=10$. }
\label{fig:6}       % Give a unique label
\end{figure}
\begin{figure}[t]
  \includegraphics[width=0.5\textwidth]{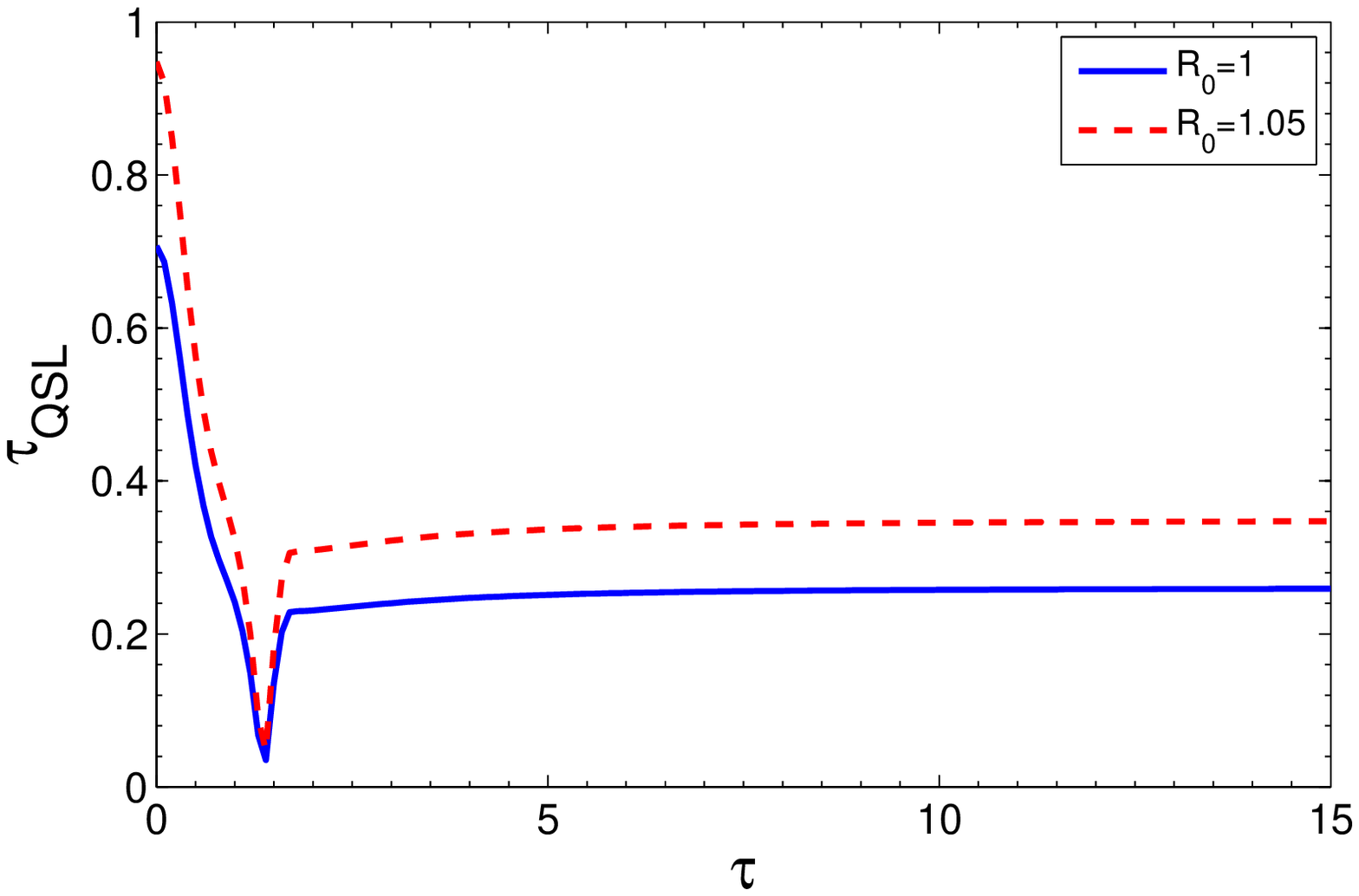}
\caption{The QSL-time  for the Ohmic-like dephasing model as a
function of the initial time parameter $\tau$ for super-Ohmic environment. The Ohmicity parameter is $s=4.5$ and $\Omega=10$.  }
\label{fig:7}       % Give a unique label
\end{figure}
\begin{figure}[t]
  \includegraphics[width=0.5\textwidth]{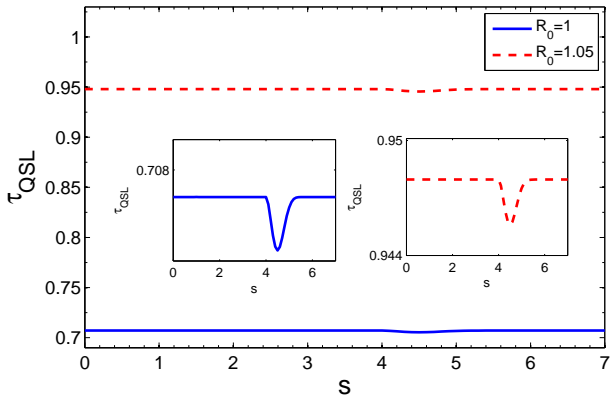}
\caption{The QSL-time  for the Ohmic-like dephasing model as a
function of the Ohmicity parameter $s$. The value of driving time is $\tau_{D}=1$, initial time is $\tau=0$ and $\Omega=10$.}
\label{fig:8}       % Give a unique label
\end{figure}
\begin{figure}[t]
  \includegraphics[width=0.5\textwidth]{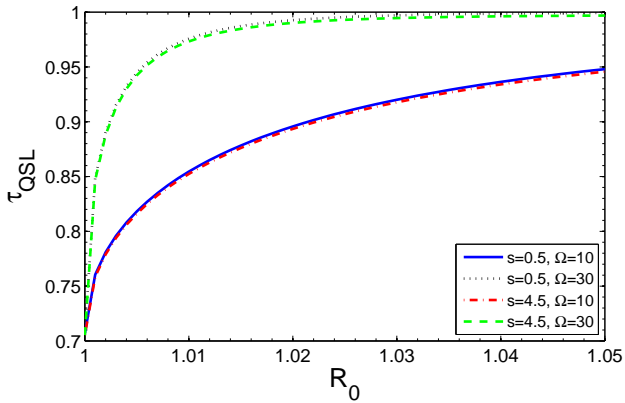}
\caption{The QSL-time  for the Ohmic-like dephasing model as a
function of the  $R_{0}$. The value of driving time is $\tau_{D}=1$ and initial time $\tau=0$. }
\label{fig:9}       % Give a unique label
\end{figure}

In Fig.(\ref{fig:6}) the QSL-time is plotted  as a function   of the initial time parameter $\tau$, when Ohmicity parameter is $s=0.5$ and the dynamic is Markovian . As can be seen QSL-time is increased by increasing the distance $R_{0}$, between Bob and the event horizon. It is Due to the fact that the QSL-time for Ohmic-like dephasing model is depend on quantum coherence in linear manner, i.e. the QSL-time is increased(decreased) by increasing(decreasing) quantum coherence \cite{Zhang}. Fig. (\ref{fig:7}) represents the QSL-time as a function   of the initial time parameter $\tau$ , when Ohmicity parameter is $s=4.5$ and the evolution is Non-Markovian\cite{Breuerx}. As can be seen in Fig.(\ref{fig:7}), there exist a protuberance phenomenon for QSL-time. This protuberance occurs due to non-Markovianity of the dynamics. In non-Markovian  dynamics in some time intervals information flow back from environment to system and it leads to the occurrence of this phenomenon for QSL-time for dephasing model. Comparison of Fig.(\ref{fig:6}) with Fig.(\ref{fig:7}) shows that for non-Markovian evolution the QSL-time is shorter than Markovian evolution. So, the speed of the evolution for non-markovian dynamic is greater than Markovian dynamic. Also one can see in both Markovian and non-Markovian evolution the QSL-time is increased by increasing $R_{0}$.

In order to show the effects of  Ohmicity parameter on QSL-time, the  QSL-time is plotted as a function of $s$ in Fig. (\ref{fig:8}) for the constant driving time $\tau_D=1$. As can be seen for all the Ohmic-like environments with different Ohmicity parameter the QSL-time is increased by increasing the distance of open quantum system from event horizon. In other words, for all Ohmic-like environments with different Ohmicity parameter , the speed of the evolution is slowed down by increasing the distance from event horizon. As can be seen there exist protuberance phenomenon for QSL-time at both $R_{0}=1$ and $R_{0}=1.05$, this protuberance phenomenon stems from non-Markovianity of the quantum evolution for some intervals of Ohmicity Parameter.  In order to show the effect of the $\Omega$(or Hawking temperature $T_H$) and $R_0$ on QSL-time, in Fig. (\ref{fig:9}) the QSL-time is plotted as a function of $R_{0}$ for Markovian ($s=0.5$), and non-Markovian ($s=4.5$) evolution for different value of $\Omega$ and Ohmicity parameter $s$. As can be seen for both of Markovian and non-Markovian evolution the QSL-time is decreased by decreasing $R_{0}$. Also, this figure shows that the QSL-time is increased by increasing $\Omega$. 

In summary, the important results about QSL-time in  Schwarzschild space-time for Ohmic-like dephasing  model are given as follow : First, QSL-time increases when the distance of Bob from event horizon, i.e. $R_0$ increases. In other word the speed of quantum evolution in Ohmic-like dephasing model decreases when the distance of Bob from event horizon increases. Second, QSL-time for non-Markovian regime (strong coupling) is shorter than Markovian regime (weak coupling). Third, QSL-time increases when the mode frequency $\Omega$ increases. Due to the reverse dependency of $\Omega$ on Hawking temperature $T_H$, it is right to say that QSL-time is increased by decreasing Hawking temperature. 
\section{Conclusion}\label{sec:5}

In this work  QSL-time in Schwarzschild space-time has been studied for two types of decoherence model. The results obtained in this article are in satisfactory agreement  to those obtained previously by Zhang et al. for damped Jaynes-Cummings and Ohmic-like dephaing models   \cite{Zhang}. In this work, it is showed that for damped Jaynes-Cummings model the QSL-time has the inverse relation with quantum coherence in Schwarzschild space-time. The quantum coherence is decreased by decreasing the distance between of quantum state with the event horizon, thus the QSL-time for damped Jaynes-Cummings model   in Schwarzschild space-time will be increase  when $R_{0}$ decreases.  The situation for QSL-time in Schwarzschild space-time and Ohmic-like dephaing model  is completely different.   In this model the QSL-time is depend on quantum coherence linearly\cite{Zhang}. It is observed that in Schwarzschild space-time for Ohmic-like dephasing model, the QSL-time is increased by increasing $R_{0}$ and $\Omega$. The results of this study may be useful to explore the speed of evolution of more complex quantum systems from the standpoint of relativity that can be used in quantum information processing in the presence of quantum noise.

%\begin{table}
%\begin{tabular}{lll}
%\hline\noalign{\smallskip}
%\end{tabular}
%\end{table}

%\begin{acknowledgements}
%\end{acknowledgements}



\end{document}